\documentclass[journal]{IEEEtran}
\usepackage{algorithmic}
\usepackage{algorithm}
\makeatletter
\newcommand{\removelatexerror}{\let\@latex@error\@gobble}
\makeatother

\usepackage{ifpdf}
\ifpdf
\else
\fi
\usepackage{cite}

\ifCLASSINFOpdf
\usepackage[pdftex]{graphicx}
\graphicspath{{../pdf/}{../jpeg/}}
\DeclareGraphicsExtensions{.pdf,.jpeg,.png}
\else
\usepackage[dvips]{graphicx}
\graphicspath{{../eps/}}
\DeclareGraphicsExtensions{.eps}
\fi
\usepackage{amsmath}
\usepackage{algorithmic}
\usepackage{array}
\ifCLASSOPTIONcompsoc
\usepackage[caption=false,font=normalsize,labelfont=sf,textfont=sf]{subfig}
\else
\usepackage[caption=false,font=footnotesize]{subfig}
\fi
\usepackage{fixltx2e}
\usepackage{stfloats}
\ifCLASSOPTIONcaptionsoff
\usepackage[nomarkers]{endfloat}
\let\MYoriglatexcaption\caption
\renewcommand{\caption}[2][\relax]{\MYoriglatexcaption[#2]{#2}}
\fi
\usepackage{url}
\begin{document}
	\title{Memory leak detection algorithm Based on Defect Mode in smart grid}
	
	\author{Ling~Yuan,
		Siyuan~Zhou,
		Neal~N.~Xiong ,~\IEEEmembership{Senior~Member,~IEEE}% <-this % stops a space
		\thanks{Ling Yuan is with the Department of Computer Science, Huazhong University of Science and Technology. Corresponding author: cherryyuanling@hust.edu.cn}% <-this % stops a space
		\thanks{Siyuan Zhou is with Huazhong University of Science and Technology.}
		\thanks{Neal N. Xiong is with College of Intelligence and Computing, Tianjin University. Corresponding author: dnxiong@126.com.}}% <-this % stops a space
	\maketitle

	\begin{abstract}
		With the expansion of the software scale and complexity of smart grid systems, the detection of power grid software defects has become a research hotspot and difficult problem.Because of the large scale of the existing power grid software code, the efficiency and accuracy of the existing power grid defect detection algorithms are not high. This paper presents a static memory leak detection method based on a defect mode. From the viewpoint of the analysis of existing memory leak defect modes, this paper summarizes memory operation behaviors (allocation, release and transfer) and presents a state machine model that is based on memory operation behavior as the foundation of static detection in memory leaks. Our proposed method employs a fuzzy matching algorithm that is based on regular expression to determine the memory operation behaviors that exist in the code to be detected and then analyzes the change in the state machine to assess the vulnerability in the source code. To improve the efficiency of detection and solve the problem of repeated detection at the function call point, we propose a function summary method for memory operation behaviors. The experimental results demonstrate that the method we proposed has high detection speed and accuracy. The algorithm we proposed can identify the defects of the smart grid operation software and ensure the safe operation of the grid.
	\end{abstract}
	

	% \fi
	\IEEEpeerreviewmaketitle

	\section{Introduction}
	\IEEEPARstart{W}{ith} With the widespread use of computer technology in smart grid operations, the scale of smart grid software is getting bigger and bigger. Therefore, defects in the smart grid software are inevitable. These defects can be fatal and are related to the safe operation of smart grid software. Among the multitudinous defects, the occurrence of memory leaks is fatal to a continuously running software program such as smart grids software. The sustained occurrence of memory leaks will cause a sharp decrease in the available memory of a server, which will affect the system capability and even cause server downtime,Which led to the collapse of the entire smart grid. Therefore, memory leak detection algorithms are urgently needed to detect defects in the entire system. Memory leak detection is generally divided into static detection and dynamic detection. Dynamic detection identifies related leaks based on dynamic changes of a program during its execution. The results of dynamic detection are accurate but the execution path depends on test cases, and the finiteness of the test cases will produce an excessive false negative rate. Dynamic detection will usually detect a program using instrumentation, which may cause defects and increase the invasiveness of dynamic detection. Static detection enables the detection of potential defects by analyzing binary source code without executing the code and can efficiently improve the software correctness and help software debuggers efficiently detect potential memory leak defects in a program. Compared with dynamic detection, static detection has some advantages: it can detect defects in the early stage of program development and reduce the cost of maintenance in subsequent stages. Test cases do not need to be artificially designed and input, which solves the software testing limitation caused by incomplete test cases. However, this method also has the disadvantages of slow detection speed and serious false positives. Therefore, reducing the false positive rate of memory leak detection to improve the detection efficiency is the focus of this paper, which has research significance.
	
	The contributions of this paper are described as follows:
	
	1. Summarize and analyze the common memory leak defect modes in C/C++ and design a fuzzy matching algorithm based on regular expression. In this paper, the summarized defect modes are analyzed by simple lexical analysis, and then the lexical units are classified into variables, keywords and numbers to form a series of defect strings in the form of regular expressions to be matched. Auxiliary methods such as path sensitivity analysis, interval arithmetic and alias analysis are employed to improve the accuracy of memory leak detection and reduce the false positive and false negative rates.
	
	2. Summarize memory operation behaviors based on the memory leak defect mode and divide them into three types: allocation, deallocation and transfer. These three types of memory operation behaviors are abstractly described, and the state machine is designed. The possible memory operation behaviors in the program can be detected using the fuzzy matching algorithm based on regular expression to control the state change of the state machine model-based memory operation behavior to detect the memory leak defect.
	
	3. To improve the efficiency of detection and reduce the scan times of program functions, a function summary method for memory operation behaviors is proposed to achieve a brief description of memory operation behaviors in functions. Static detection methods based on a state machine are divided into a general method and a special method. General memory leak detection is combined with a function summary and a control flow graph (CFG) based on the state machine to achieve memory leak detection. The special detection is static detection based on rules.
	
	4. The experimental results demonstrate that the method designed in this paper has higher detection speed and accuracy.The algorithm we proposed can be widely used in defect detection of major smart grid software.
	
	The paper is organized as follows: In Section 2, we present related studies. Section 3 discusses our preliminaries. Section 4 discusses  the proposed algorithm for memory leak detection. In Section 5, we present experiments simulation and results analysis are illustrated. In Section 6, we discusses the study’s conclusions.

	%子标题
	%\subsection{Subsection Heading Here}
	%Subsection text here.
	%
	%% needed in second column of first page if using \IEEEpubid
	%%\IEEEpubidadjcol
	%
	%\subsubsection{Subsubsection Heading Here}
	%Subsubsection text here.
	
	\section{RELATED WORK}
	For memory leak detection, dynamic detection \cite{clause2010leakpoint}\cite{nethercote2007valgrind}\cite{reed1991purify} involves executing test cases and identifying potential memory leak defects in the program to be detected, such as LeakPoint \cite{clause2010leakpoint}, valgrind \cite{nethercote2007valgrind}, Purify \cite{reed1991purify}, Maebe \cite{cherem2007practical}, and SafeMem \cite{qin2005safemem}. Dynamic detection has the advantages of a low false positive rate and a short detection time; its disadvantages include an overly high false negative rate and excessive dependence on the inputted use cases.
	
	The static detection of a memory leak utilizes static analysis technology (symbolic execution, context sensitivity analysis, and function summary generation) to detect potential memory leak defects in a program without executing the program. In recent years, considerable research on the static detection of memory leaks has been performed.
	
	Heine and Lam \cite{heine2003practical} proposed a memory detection model for C/C++ based on flow sensitivity and context sensitivity: Clouseau. In the model, the memory object can only be owned by one pointer. Sensitivity is used to analyze the transfer of pointer ownership to detect a memory leak. YuLei and Ding et al. \cite{sui2012static}\cite{tan2017efficient}\cite{yan2016automated}\cite{sui2014detecting} designed a memory leak detection tool based on value-flow analysis: SABER. The tool abstracts the program into a fully sparse value-flow graph, calculates the data flow information on the node and analyzes the accessibility of nodes on the graph to determine the existence of a memory leak defect. SATURN \cite{xie2005context} is a memory leak detection method that is based on a function summary proposed by Xie Yichen and Alex Aiken et al. This method records the boolean value that is used to describe the memory operation in the function summary and other related information and implements the memory leak detection among procedures by combining path and context sensitivity analysis. However, the nonprocessing of function parameters may cause imprecision in the set of memory space ownership variables while analyzing procedures. Similarly, the memory leak detection tool SPARROW, proposed by Jung and Yi \cite{jung2008practical} is a type of path-insensitive analysis. The tool abstracts functions into parameterized expressions of function summaries, applies them to corresponding function call statements, and analyzes the correlation of the return values. Xu \cite{xu2015melton} proposed the Memory State Transition Graph (MSTG) and implemented the tool Melton. MSTG recorded the change in the memory-object states as path conditions. Cherem [4] converted the problem into a reachability problem and analyzed the numerical spread on the dataflow graph to detect memory leaks.
	
	\section{PRELIMINARIES}
	\subsection{SOLUTION OVERVIEW}
\begin{figure}[!t]
		\centering
		\includegraphics[width=2.5in]{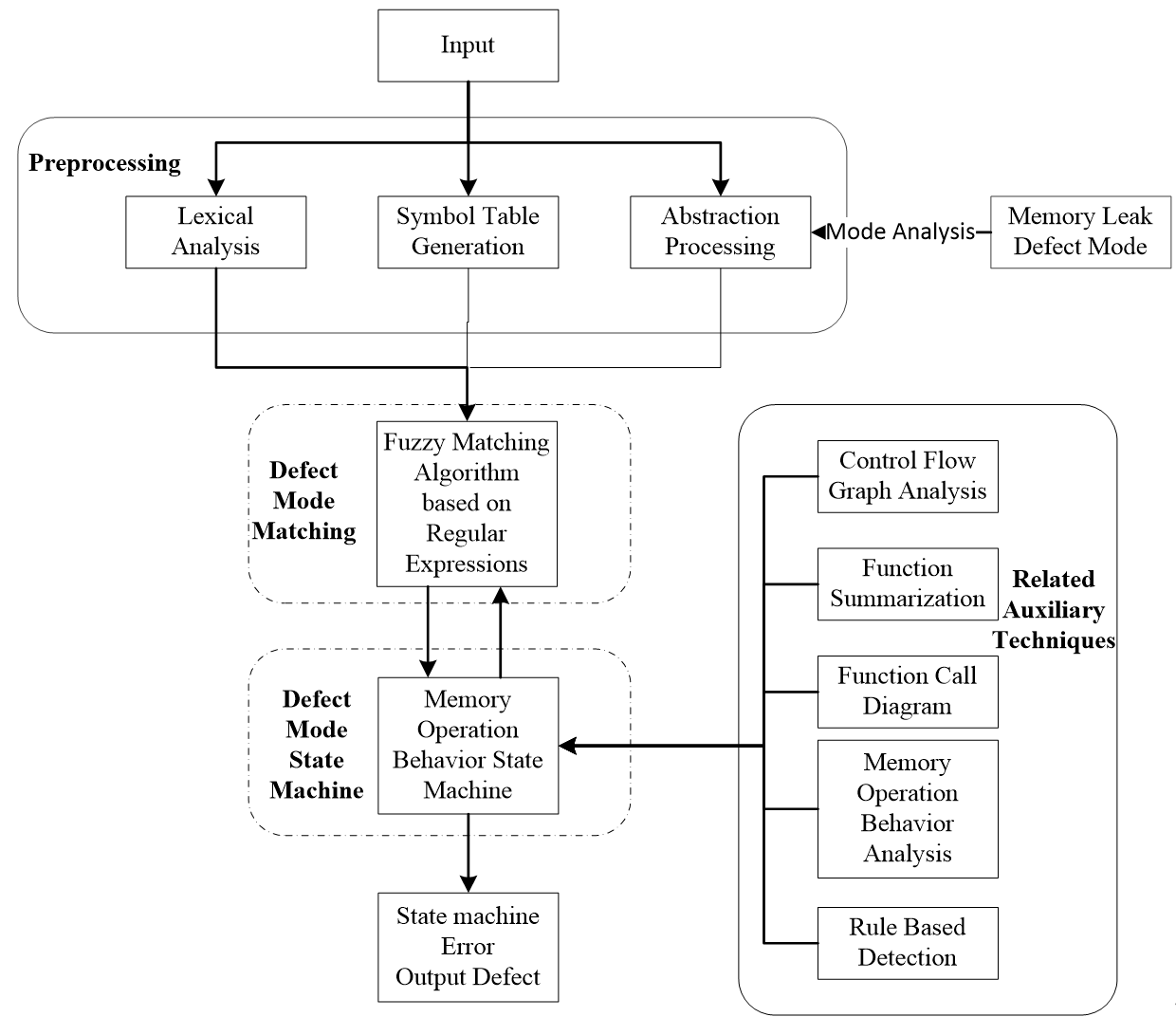}
		\DeclareGraphicsExtensions.
		\caption{Static detection architecture diagram based on defect mode}
		\label{Static detection architecture diagram based on defect mode}
	\end{figure}

	Our proposed static memory leak detection method is based on the defect mode, applies its matching algorithm to obtain potential defect statements in the source code to be detected, and combines relevant auxiliary techniques to control state changes in the Defect Mode State Machine. We can find potential defects by analyzing the current state information, as shown in Fig. \ref{Static detection architecture diagram based on defect mode}.
	
	As shown in Fig. 1, the core of static detection methods for a memory leak based on the defect mode is detailed as follows:
	
	1. Defect Mode Matching: Using a matching algorithm to obtain the defect points (memory operation behavior) that can be used to control the state change of the state machine from the program to be detected.
	
	2. Defect Mode State Machine: Using the matched defect  points to control the state change in the Defect Mode State Machine to obtain potential defects in the program.
	
	3. Auxiliary Techniques: This part employs related techniques with State Machine to the detection of the defects, which primarily includes the function summary and CFG.
	
	\subsection{MEMORY LEAK DEFECT MODE ANALYSIS}
	
	The software defect mode \cite{quinlan2007techniques} is a summary of certain types of defects or errors that often occur in a program, including the forms of the defects and the conditions of their generation. Common memory leak defect modes \cite{quinlan2007techniques} include missing release memory leak, pointer memory leak, memory leak of mismatched request and release, and class member memory leak.
	
	\subsubsection{MISSING RELEASE MEMORY LEAK}
	
	This type only includes memory leaks caused by local variables within a function, excluding interprocedure memory leaks caused by global variables and function calls. This error is the most common and easily identified error in memory leaks, including the following two types.
	
	1. The normal missing release
	
\begin{figure}[!t]
		\centering
		\includegraphics[width=2.5in]{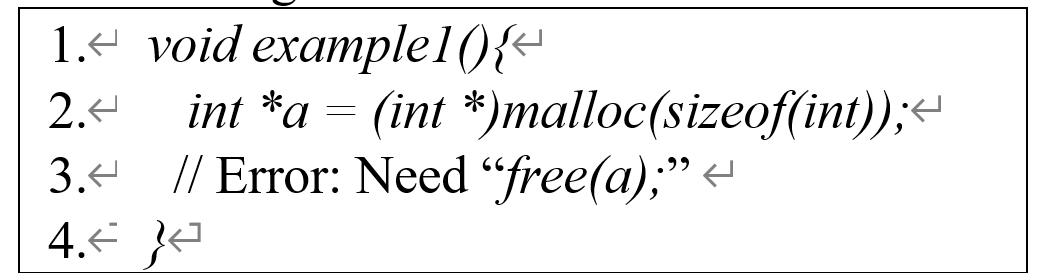}
		\DeclareGraphicsExtensions.
		\caption{An example of the normal missing release}
		\label{An example of the normal missing release}
	\end{figure}

	It refers to that memory that is applied in a function but not returned to the system in time. The cause of this defect is relatively simple, that is, no reasonable release of memory causes the memory leak. An example of this type of memory leak is shown as Fig. \ref{An example of the normal missing release}.
	
	2. The path-sensitive missing release
	
\begin{figure}[!t]
		\centering
		\includegraphics[width=2.5in]{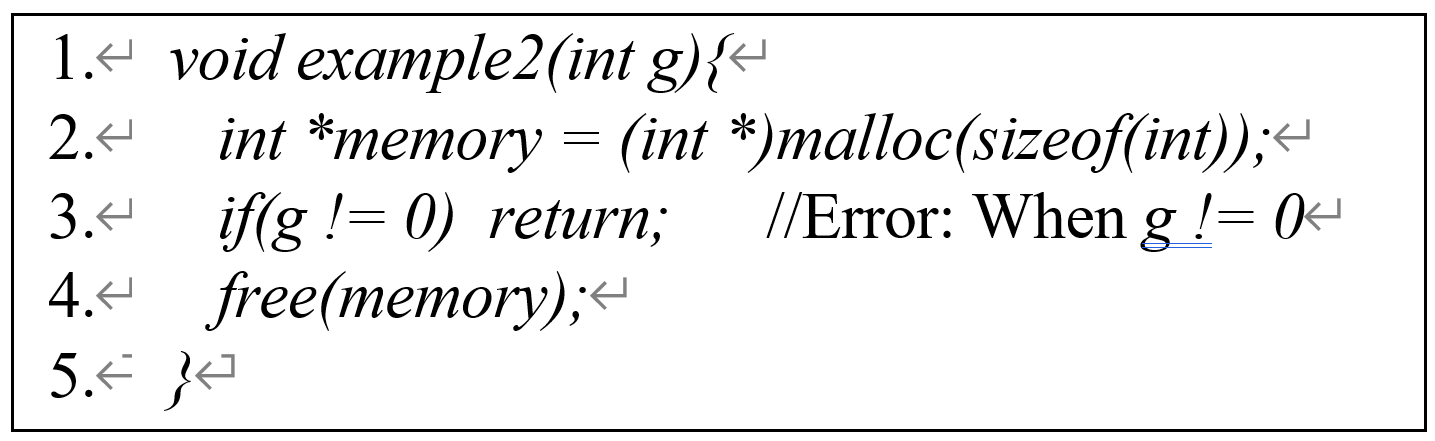}
		\DeclareGraphicsExtensions.
		\caption{An example of the path-sensitive unreleased}
		\label{An example of the path-sensitive unreleased}
	\end{figure}

	The program does not release allocated memory before all paths are ended. This type of memory leak is caused by the difference in the execution paths due to conditional statements, loops and other causes and excessively complex paths or function return in advance in the execution process of a function. In this case, the correct memory request and release statements are usually included without being completely executed because the function may have different execution paths due to different inputs during each execution. An example of the path-sensitive unreleased is shown as Fig. \ref{An example of the path-sensitive unreleased} .
	
	\subsubsection{POINTER MEMORY LEAK}
	
\begin{figure}[!t]
		\centering
		\includegraphics[width=2.5in]{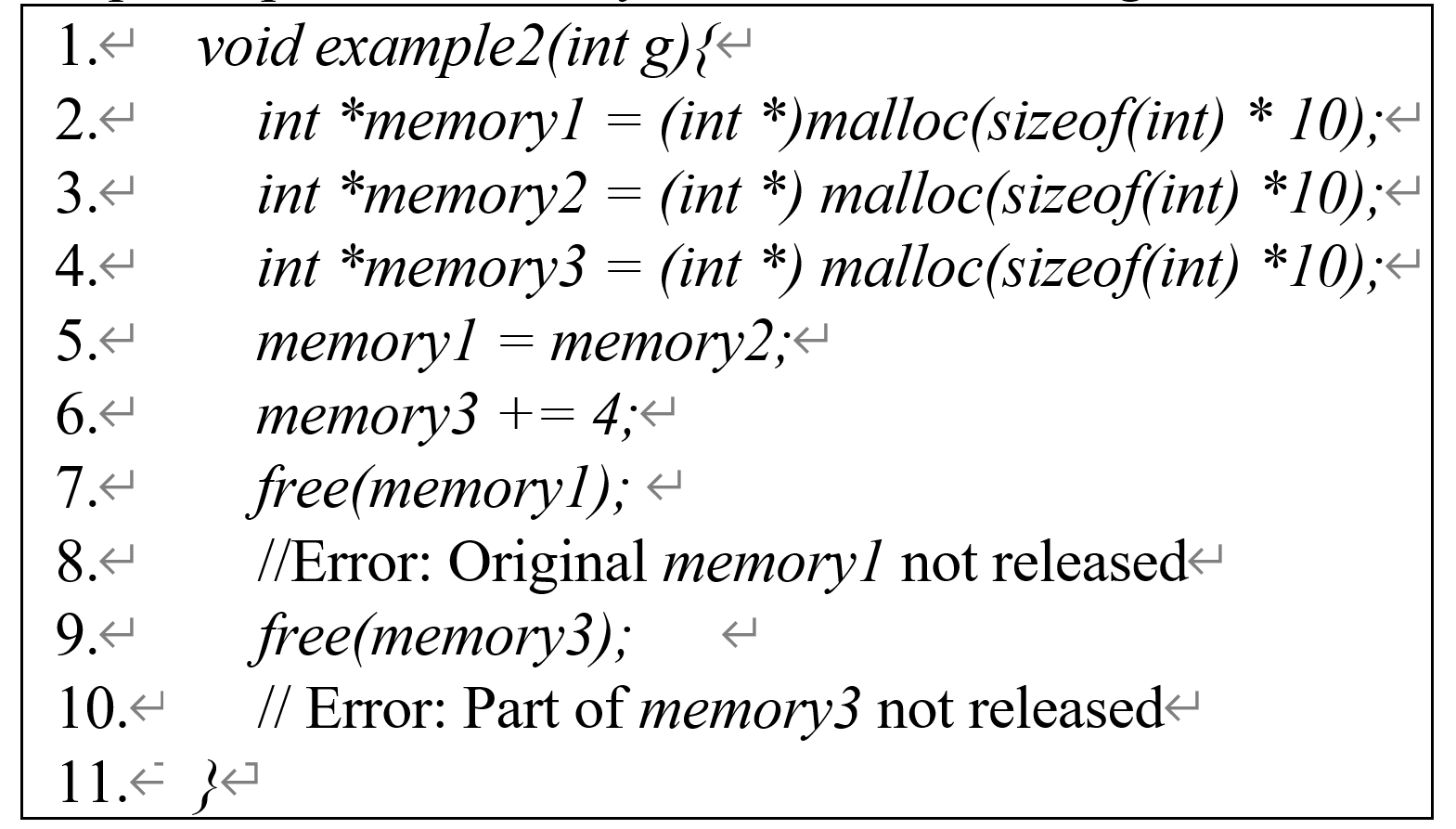}
		\DeclareGraphicsExtensions.
		\caption{An example of pointer memory leak}
		\label{An example of pointer memory leak}
	\end{figure}

	A pointer memory leak refers to an incomplete or erroneous release of memory caused by an incorrect operation of the pointer that owns the ownership of memory, which causes a loss of part or full ownership. This type of defect is primarily caused by an unreasonable operation of a pointer, such as illegal operations of a pointer and pointer loss or redirection in the process of a function call. An example of pointer memory leak is shown as Fig. \ref{An example of pointer memory leak} .
	
	\subsubsection{MEMORY LEAK OF MISMATCHED ALLOCATION AND RELEASE}
	
	This type refers to the inconsistency of the functions used to request and release memory. For example: “$new[]$” is used to request space, whereas “$delete$” is used to free memory instead of “$delete[]$”, and the memory space requested by “$malloc$” is released by deleting. This type of error is easy to find in static detection and is often caused by the negligence of programmers.
	
	\subsubsection{CLASS MEMBER MEMORY LEAK}
	
	A class member memory leak refers to defects caused by the incorrect operation of class members. This type of leak includes the following three types.
	
	1. Copy Constructor/Assignment Overloaded Function 
	
\begin{figure}[!t]
		\centering
		\includegraphics[width=2.5in]{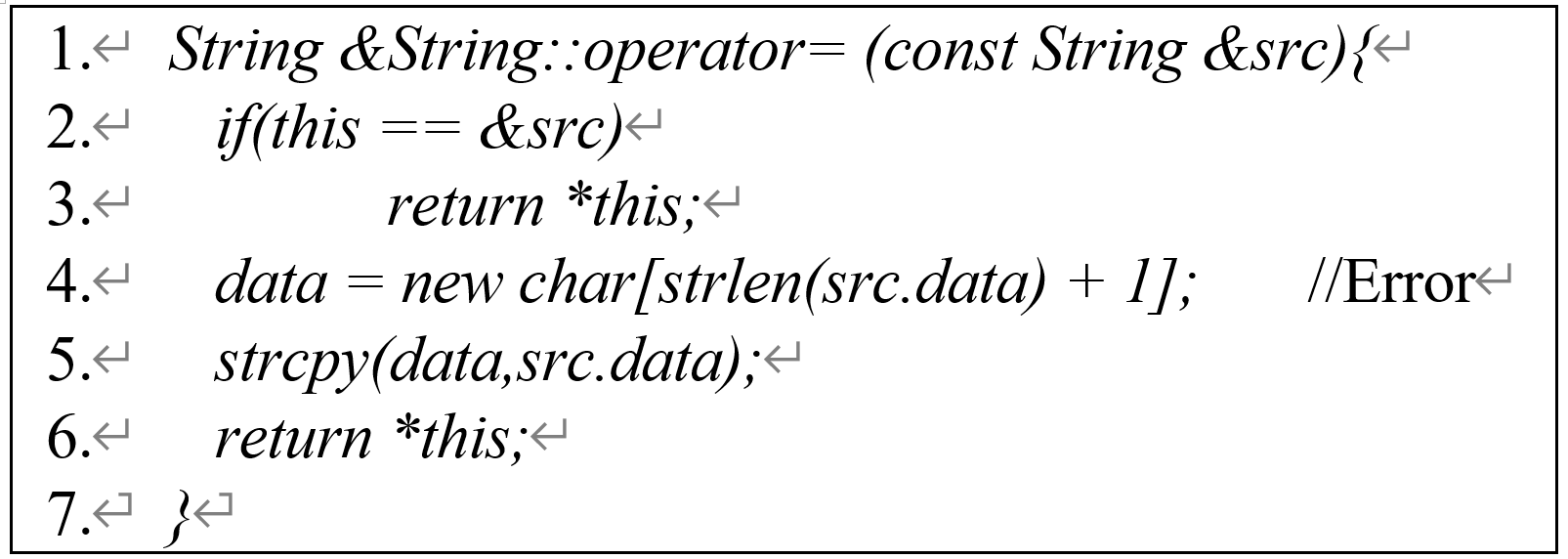}
		\DeclareGraphicsExtensions.
		\caption{An example of incorrect assignment overloaded function}
		\label{An example of incorrect assignment overloaded function}
	\end{figure}

	For this type of error, a pointer object exists in class members. When the definition of a copy constructor or assignment overloaded function is incorrect, a memory leak may occur. When the copy constructor/assignment overloaded function is undefined but the class members include a pointer, the compiler will generate the two functions by default. The default function generated by the compiler will cause repeated application and release of memory and even shallow copy defects. An example of incorrect assignment overloaded function is shown as Fig. \ref{An example of incorrect assignment overloaded function} .
	
	2. Constructor/Destructor Memory Leak
	
\begin{figure}[!t]
		\centering
		\includegraphics[width=2.5in]{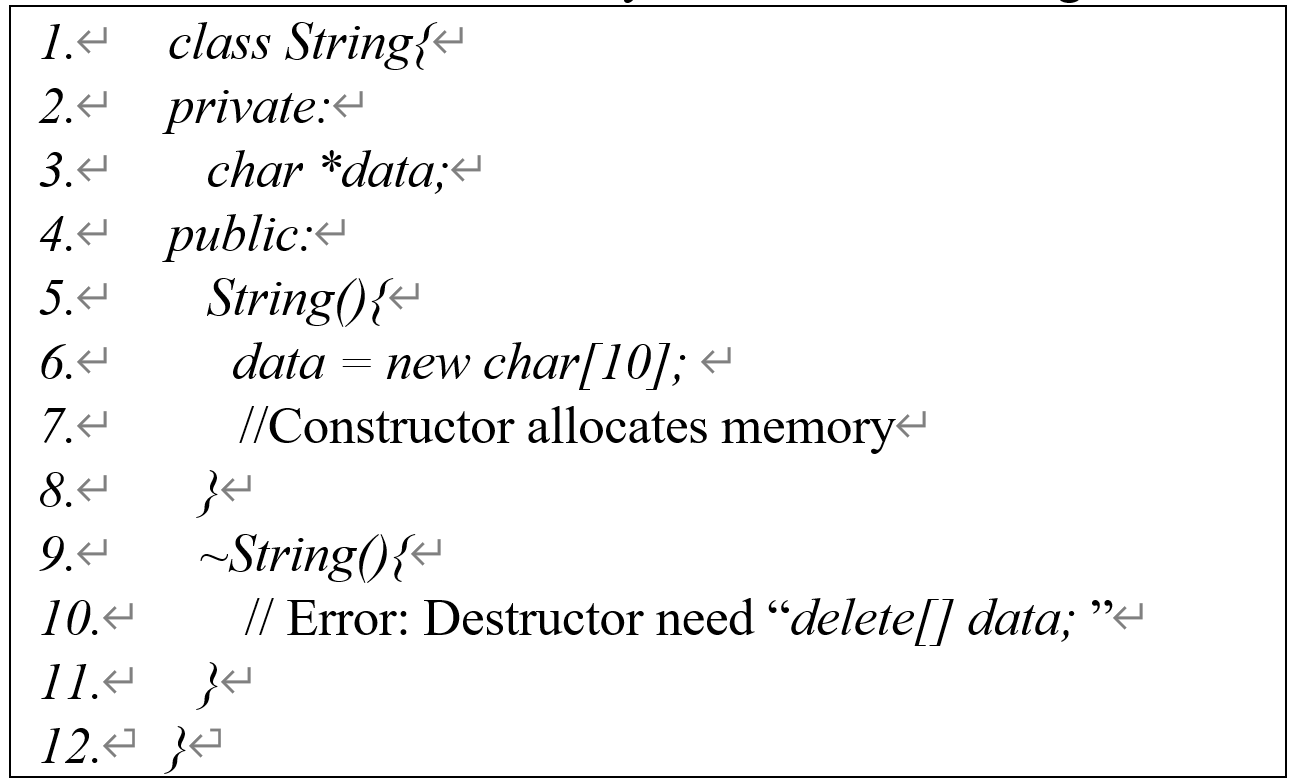}
		\DeclareGraphicsExtensions.
		\caption{An example of constructor destructor memory leak}
		\label{An example of constructor destructor memory leak}
	\end{figure}

	This kind of leaks happens when a class contains pointers for which the memory operation behavior is inconsistent in the constructor and destructor. After the memory is allocated, the memory space is not reasonably released in the destructor, which causes a memory leak. An example of constructor/destructor memory leak is shown as Fig. \ref{An example of constructor destructor memory leak} .
	
	3. Inheritance Relational Memory Leak
	
\begin{figure}[!t]
		\centering
		\includegraphics[width=2.5in]{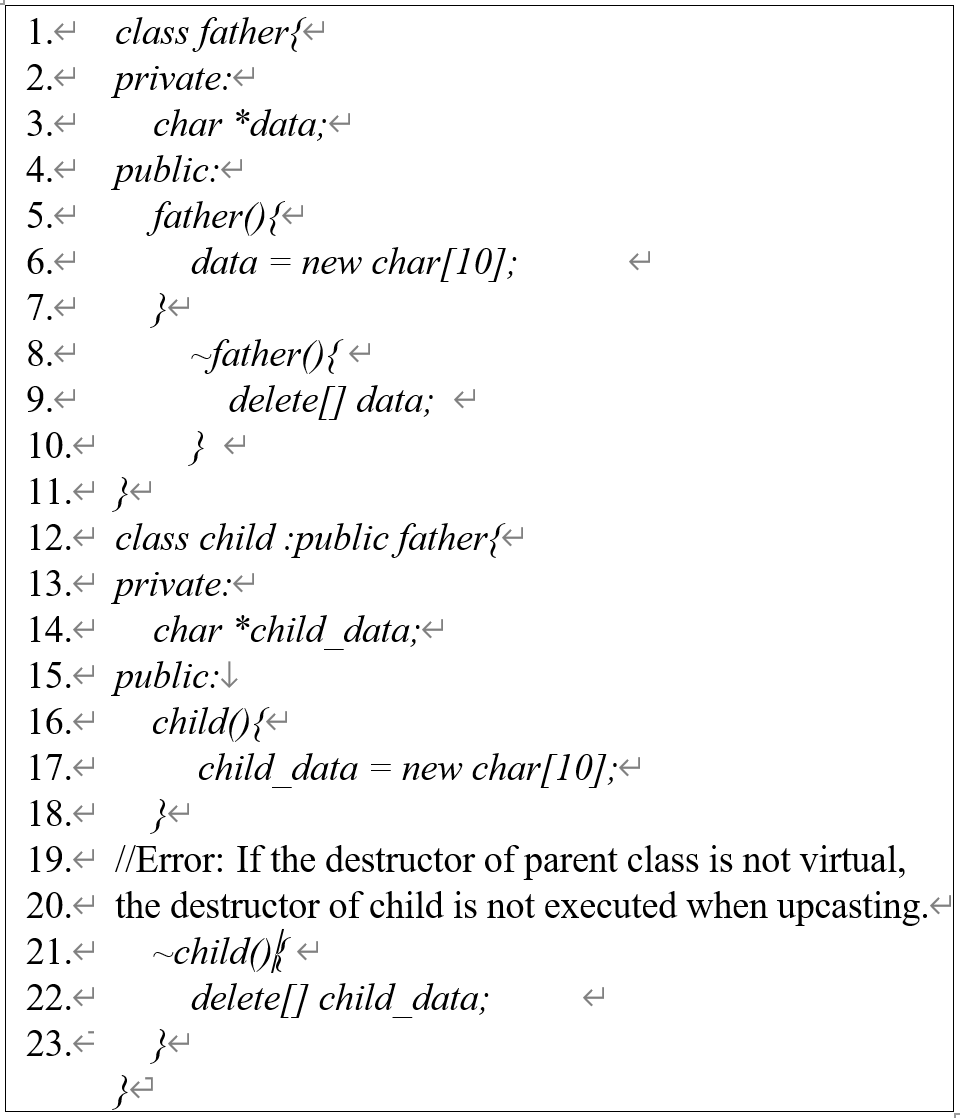}
		\DeclareGraphicsExtensions.
		\caption{An example of inheritance relational memory leak}
		\label{An example of inheritance relational memory leak}
	\end{figure}

	This type of error primarily occurs between parent classes and child classes. In the case in which the original memory space is not released, the subclass reapplies space for the member variable in the parent class, which prevents the original space from being released; The destructor of the parent class is nonvirtual. When going up, the parent class cannot use the polymorphic mechanism to reasonably free the memory space requested in the subclass. An example of inheritance relational memory leak is shown as Fig. \ref{An example of inheritance relational memory leak} .
	
	\subsection{PREPROCESSING}
	
	For memory leak detection based on the defect mode, first, the program to be analyzed needs to be preprocessed by lexical analysis, and a doubly linked list of lexical unit is generated. Second, a symbol table for the entire file is created by interval division, and the lexical unit node attribute is combined with its scope to establish a symbol table in the scope.
	
	\subsubsection{THE LEXICAL UNIT}
	
\begin{figure}[!t]
		\centering
		\includegraphics[width=2.5in]{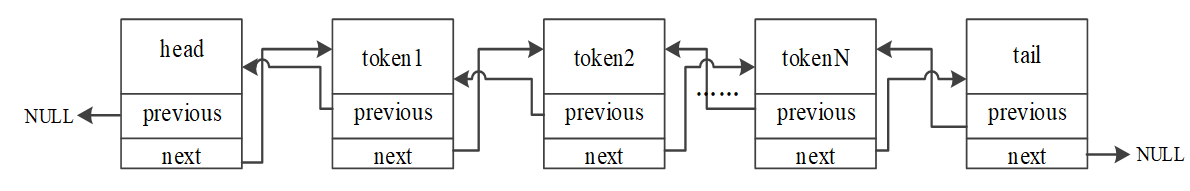}
		\DeclareGraphicsExtensions.
		\caption{Doubly linked list for lexical units}
		\label{Doubly linked list for lexical units}
	\end{figure}

	The lexical unit \cite{cui2010code} is the smallest description of a single keyword in the source code. The attribute information of a single lexical unit, is recorded. The lexical unit chain is simultaneously constructed using the predecessor pointer and the backward pointer, and a doubly linked list of the entire program is established for memory leak detection. The structure of the doubly linked list is shown in Fig. \ref{Doubly linked list for lexical units} .
	
	\subsubsection{SYMBOL TABLE}
	
\begin{figure}[!t]
		\centering
		\includegraphics[width=2.5in]{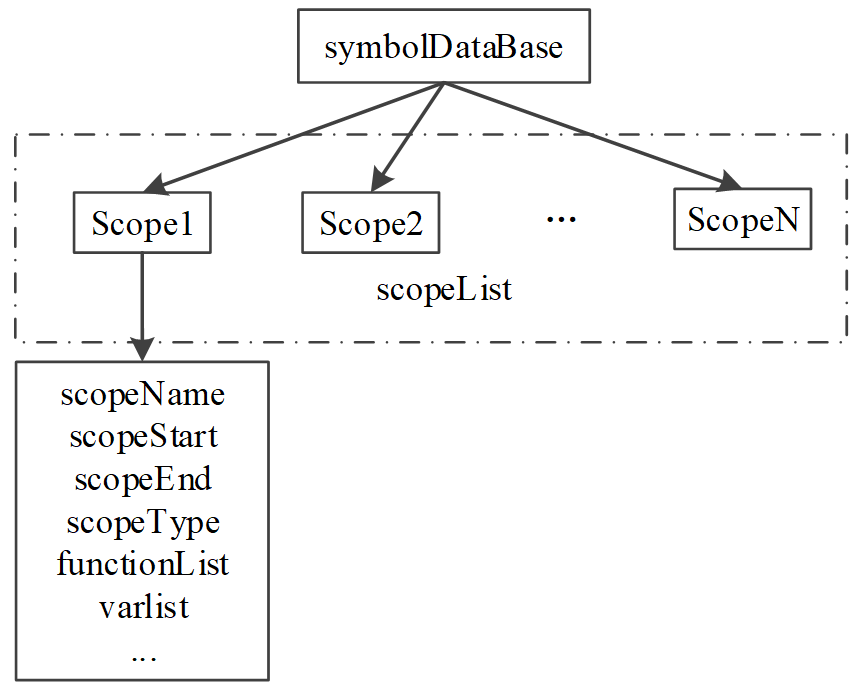}
		\DeclareGraphicsExtensions.
		\caption{Structure of the symbol table}
		\label{Structure of the symbol table}
	\end{figure}

	A symbol table \cite{khurshid2003generalized}\cite{li2010practical}\cite{king1976symbolic} is used to record the program scope and related information within the scope, which is a supplement to the lexical unit attributes. This table represents the mapping relationship between the lexical unit and the the scope and is built into a tree. A tree-shaped symbol table structure is suitable for traversal, search, and layering. The structure design of the symbol table is shown in Fig. \ref{Structure of the symbol table} .
	
	The scopes in the symbol table include the global domain ($eGlobal$), class domain ($eClass$), structure domain ($eStruct$), union domain ($eUnion$), namespace ($eNamespace$), function domain ($eFunction$), if statement block domain ($eIf$), and so on.
	
	\subsubsection{ DEFECT PATTERN ABSTRACTION PROCESSING}
	
	For the defect mode, the pattern string of the defect mode should be general. The programs are different due to the different structures of different programs. Matching the general defect patterns with the special programs is a problem that needs to be solved in the defect pattern matching process.

	\begin{table}[!t]
		\renewcommand{\arraystretch}{1.3}
		\caption{EXPLANATION OF ABSTRACTION OF DEFECT PATTERN}
		\label{EXPLANATION OF ABSTRACTION OF DEFECT PATTERN}
		\centering
		\begin{tabular}{c p{5.5cm}}
		\hline\hline
		Symbol    & Meaning                                                                                          \\
		\hline
		\%any\%   & Any form of   keywords, which can be any lexical units, such as variables, types, and operators. \\
		\%name\%  & A variable name   or type. For example, “int a” can   be expressed as “\%name\% \%name\%”.       \\
		\%type\%  & A variable type.   For example, “int a” can be   expressed as “\%type\% \%name\%”.               \\
		\%num\%   & A number, such as   “23”.                                                                        \\
		\%bool\%  & A Boolean value,   “true” or “false”                                                             \\
		\%comp\%  & A comparison   operator, such as “\textgreater{}”, “==”, etc.                                    \\
		\%str\%   & A string.                                                                                        \\
		\%var\%   & A variable.                                                                                      \\
		\%varid\% & A variable ID.                                                                                   \\
		\%or\%    & “$|$”                                                                                              \\
		\%oror\%  & “$||$”                                                                                             \\
		\%op\%    & An operator, such   as “=”.    \\   
		\hline\hline                                  
		\end{tabular}
	\end{table}
	
	Using a summary of C/C++, the keyword categories in the program, including variables, numbers, and operators, can be statistically analyzed. A summary of the keywords in the defect mode is provided in Table \ref{EXPLANATION OF ABSTRACTION OF DEFECT PATTERN} .
	
	\subsection{MEMORY OPERATION BEHAVIORS ANALYSIS}
	
	The analysis of the memory operation behaviors is helpful for establishing a state machine based on memory operation behaviors. And then we can analyze state transition of the state machine in the program to determine whether a memory leak defect exists. The memory operation behaviors include the following three types:
	
	\subsubsection{ALLOCATE}
	
	Allocation refers to the process of obtaining the memory space by the functions ($new, malloc, alloc$) provided by C/C++. The allocation of memory is the start position of memory leak detection. We use the defect pattern matching algorithm to determine whether the memory allocation exist. If the answer is yes, the state machine that belongs to this memory is established, and then the state of the state machine is analyzed to detect any memory leaks. An abstract description of the allocating memory behavior is expressed as

	\begin{equation}
	\begin{aligned}
	\text { Alloced }(p)=&\left\{\left(v_{1}(v), w_{1}\right),\left(v_{2}(v), w_{2}\right),\left(v_{3}(v), w_{3}\right)\right.\\
	& \ldots,\left(v_{n}(v), w_{n}\right)
	\end{aligned}
	\end{equation}
	
	where $Alloced(p)$ represents a collection of memory allocation behaviors that occur within the program. $\left(v_{i}(v), w_{i}\right)$ represents a memory allocation, in which $v_{i}(v)$ represents a collection of variables that have the ownership of the memory space. The variables within the collection can be local variables, global variables, and static variables. If $v_{i}(v)$ is empty, the memory may have been released or a memory leak defect has occurred. $w_{i}$ indicates the function for this memory allocation.
	
	\subsubsection{FREE}
	
	Free refers to the process of memory release by the functions ($free, delete, delete[]$) provided by C/C++. For a unique memory space, the behavior of the releasing memory is not unique. In some cases, a variable set has a memory space that releases memory multiple times; it is a defect that must be identified. An abstract description of the memory release is expressed as
	
	\begin{equation}
	\begin{aligned}
	\operatorname{Free}(p)=&\left\{\left(v_{1}(v), f_{1}, \text { record }\right),\left(v_{2}(v), f_{2}, \text { record }\right),\right.\\
	& \ldots,\left(v_{n}(v), f_{n} \text { record }\right)
	\end{aligned}
	\end{equation}
	
	where ${Free}(p)$ represents a collection of memory release behaviors that occur within the program. $\left(v_{i}(v), f_{i}, \text {record}\right)$ represents the memory release. $f_{i}$ represents the function for memory release, where $f_{i} \in\{ {free, delete},  {delete}[]\}$. A match check will be performed for $f_{i}$ and $w_{i}$ to check the correctness. The parameter $record$ indicates whether the memory space has been released. When $record = true$ , if the memory release behavior is performed again, an error will be reported. When there exists a memory allocation for $v_{i}(v)$ again after the memory is released, $record$ will be set to $false$ from $true$.
	
	\subsubsection{TRANSFER}
	
	Transfer refers to the change in the variable that has access to the memory space. This change can be addition, reduction, and transfer. The behaviors of variable assignments, function returns, pointer calculations will cause   $v_{i}(v)$ to change. Memory transfer occurs in the following situations:
	
	(a) Scope change: The scope of the local variable ends. For example, a variable is returned as a function return value, and needs to be removed from the collection $v_{i}(v)$ to which it belongs.
	
	\begin{table}[!t]
		\renewcommand{\arraystretch}{1.3}
		\caption{THE CHANGES CAUSED BY "$p = q$"}
		\label{THE CHANGES CAUSED BY "p = q"}
		\centering
		\begin{tabular}{lll}
			\hline\hline
			& $q \in{{v}}_{i}(v)$ & $q \notin{{v}}_{i}(v)$ \\
			\hline
			$p \in{{v}}_{i}(v)$ & No Change	 & Remove $q$ from $v_{i}(v)$ \\
			$p \notin v_{i}(v)$ & Remove $q$ from $v_{i}(v)$	 & No Change \\		 
			\hline\hline                                  
		\end{tabular}
	\end{table}

	(b) Variable assignment: For the assignment statement $p = q$, the changes for ${v}_{i}(v)$ are shown in Table \ref{THE CHANGES CAUSED BY "p = q"} .

	(c) Pointer calculation: When a pointer with the ownership of the memory space is calculated (such as "++", "--"), the relevant variables are removed from ${v}_{i}(v)$ .
	
	(d) Function call. Memory operations exist in the function. A specific analysis of the function summary is needed.
	
	When ${v}_{i}(v)$ is empty, additional operations in the memory space cannot be performed, which indicates that the memory cannot be released, that is, a memory leak defect exists.
	
	\section{MEMORY LEAK DETECTION Methodology}
	
	The detection method will be described in this section. Based on the previous memory leak defect mode, the corresponding state machine is established. Using the defect pattern matching algorithm, the memory operation behavior in the program is detected, and the state change of the state machine is controlled to discover the memory leak. To address the function call problem that occurs during the detection process, a function summary generation algorithm for memory operation behavior is proposed.
	
	\begin{figure}[!t]
		\label{Match($TokenList$, $Pattern$)}
		\renewcommand{\algorithmicrequire}{\textbf{Input:}}
		\renewcommand{\algorithmicensure}{\textbf{Output:}}
		\removelatexerror
		\begin{algorithm}[H]
			\caption{Match($TokenList$, $Pattern$)}
			\begin{algorithmic}[1]
				\begin{small}
					\renewcommand{\algorithmicrequire}{\textbf{Input:}}
					\renewcommand{\algorithmicensure}{\textbf{Output:}}
					\REQUIRE $TokenList$, $Pattern$
					\ENSURE   $firstP$: Record the location when the match was successful. If the match was not successful, it is $null$.\\
					$endP$: Record the next position of the end point when the match is successful, which is also the starting point of the next match.					
					\STATE $firstP = null, endP = null, set = null$ \\
					\WHILE {$token$ in $TokenList$}
					\WHILE {each $subP$ in $Pattern$}
					\STATE $curPattern = getCurPattern(subP)$ \\
					\IF{$curPattern$ is Single Selection Unit}
					\STATE $addSet(set,curPattern)$\\
					\ELSIF{$curPattern$ == Multiple Selections Unit}
					\IF{$expressions (a) \quad : \quad "[abc]"$}
					\STATE $multiPattern1(set,curPattern)$\\
					\ELSIF{$ expressions (b) \quad or \quad expressions(c) $}
					\STATE $multiPattern1(set,curPattern)$\\
					\ENDIF 
					\ENDIF 
					\IF{$compare(token,set)$}
					\IF{$firstP == null$}
					\STATE $firstP = token$\\
					\ENDIF
					\STATE $token = getNextToken(TokenList)$\\
					\ENDIF
					\STATE $subP = getNextPattern(Pattern)$\\
					\STATE $set = null$\\
					\IF{$(token == null \&\& subP == null) || (token != null \&\& subP = null)$}
					\STATE $endP = token$ \\
					\STATE $check(firstP, endP)$\\
					\ELSIF{$token == null \&\& subP != null$}
					\STATE contine;
					\ENDIF
					\STATE $subP = getFirst(Pattern)$
					\STATE $firstP = null$
					\ENDWHILE 
					\ENDWHILE
					\RETURN $firstP,endP$
				\end{small}
			\end{algorithmic}
		\end{algorithm}
	\end{figure}
	
	\subsection{DEFECT PATTERN MATCHING}
	
	This matching involves matching the abstracted defect pattern with the lexical unit doubly linked list to identify suspicious defect points, which is the basis for subsequent detection. We design a fuzzy matching algorithm based on regular expressions.
	
	Due to the different programming habits of each programmer, the different programming methods, and other elements, the lexical units with similar semantics will be written differently in the same position. The first problem that must be solved is the uncertainty of the lexical unit at a single location; thus, we introduce regular expressions. A regular expression is a fuzzy representation of multiple choices on a single lexical unit. The following expressions are employed:
	
	(a) "$[abc]$": It refers to many matching choices for a single character, which indicates that the character that appear at the current position can only be $"a", "b", or "c"$.
	
	(b) "$void|int|float|char$": It refers to many choices of a single lexical unit, which indicates that the lexical unit that appears at the current position can only be $"void", "int", "float", or "char"$.
	
	(c) "$void|int|float|char|$": It refers to many choices of a single lexical unit, which indicates that the lexical unit that appears at the current position can only be $"void", "int", "float", or "char"$ or none of these;
	
	(d) The abstraction of a single lexical unit is shown in Table \ref{EXPLANATION OF ABSTRACTION OF DEFECT PATTERN} .
	
	The fuzzy matching algorithm based on regular expression is designed as shown in Algorithm 1.
	
	The time complexity of Algorithm 1 is $O(d*n)$, where $n$ represents the length of the doubly linked list after the program is preprocessed; it is primarily determined by the size of the program. d represents the product of the number num of the matching string units and the longest length len of the units. In general, $num < 20, len < 10$. Therefore, $d < 200$.
	
	\subsection{STATE MACHINE CONSTRUCTION}
	
	\begin{figure}[!t]
		\centering
		\includegraphics[width=2.5in]{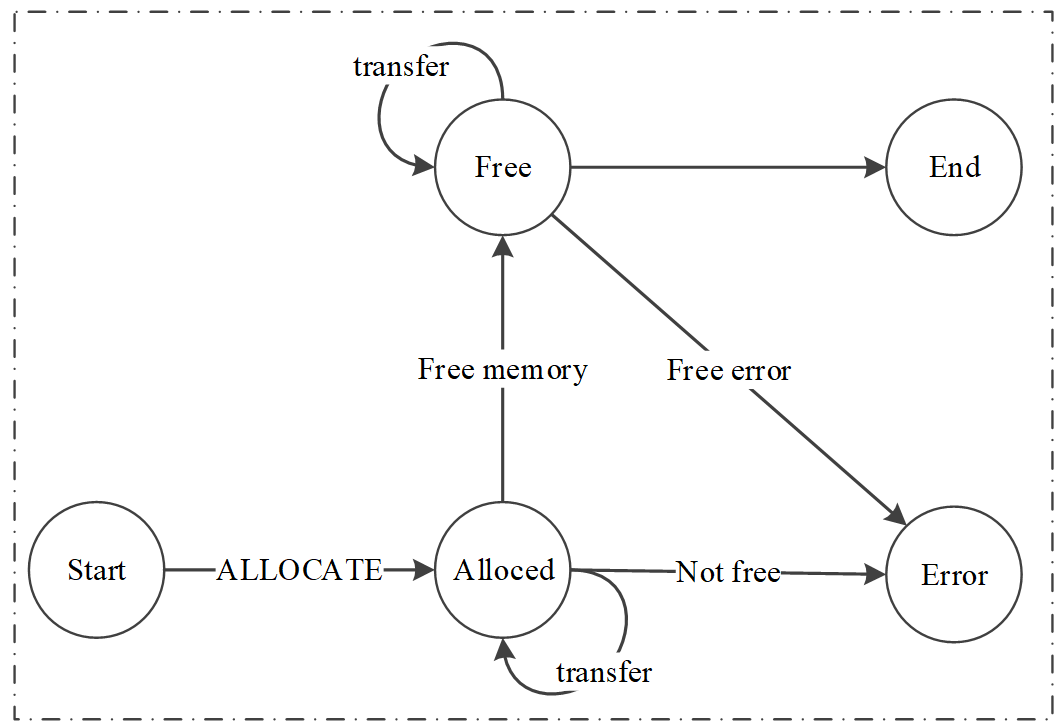}
		\DeclareGraphicsExtensions.
		\caption{ Memory operation behavior state machine}
		\label{ Memory operation behavior state machine}
	\end{figure}
	
	A state machine can be established based on the memory operation behavior. We use the proposed matching algorithm to identify the memory operation behavior in the code and analyze any related defects according to the state transition of the state machine. The memory operation behavior state machine is shown in Fig. \ref{ Memory operation behavior state machine}.
	
	\subsubsection{START → ALLOCED}
	
	When the allocation behavior (Alloc) is matched, the state of the state machine changes and enters "$Alloced$". The abstract description of this behavior is $\left(v_{i}(v), w_{i}\right)$ .
	
	\subsubsection{ALLOCED → ALLOCED: TRANSFER}
	
	After entering the state of "$Alloced$", the set of variables that have ownership of the memory space may change due to the memory transfer operations. This state change is primarily intended to record the set of variables that have changed due to the memory transfer operation.
	
	\subsubsection{ALLOCED → FREE}
	
	When the free behavior is matched, the state machine state changes from "$Alloced$" to "$Free$". The abstract description of this behavior is $\left(v_{i}(v), f_{i}, \text {record}\right)$ .
	
	\subsubsection{FREE → FREE: TRANSFER}
	
	After entering the state "$Free$", the set of variables that have ownership of the memory space may also change due to the memory transfer operation, such as pointer nulling. This state change is primarily intended to record the memory transfer situation that occurs after the memory is released.
	
	\subsubsection{FREE → END}
	
	In the case in which no error is detected, the Free state will change to "$End$" when the program detection is completed or the variables set becomes empty. This condition indicates that the memory is reasonably consumed and no defects are detected.
	
	\subsubsection{ALLOCED → ERROR}
	
	This situation indicates a potential defect, which may occur as a forced change when the program detection is completed but the state remains "$Alloced$".
	
	\subsubsection{FREE → ERROR}
	
	This situation indicates that the memory release may occur in an error and the memory release is unsuccessful. This situation may occur because the memory free function $f_{i}$ is inconsistent with the memory allocation function $W_{i}$ .
	
	\subsection{FUNCTION SUMMARY GENERATION}
	
	\subsubsection{FUNCTION SUMMARY DEFINITION}
	
	The memory operation behaviors may occur in multiple functions. For example, memory allocation is in function $a$, and the memory is freed in function $b$. Therefore, the memory operation behaviors across functions must be analyzed during the detection process. If the function expansion detection method is used for the function call point during the detection process, it may cause a repeated scan of functions, which is an extremely inefficient method. To solve this problem, a function summary technique is introduced to memory leak detection. The function summary describes the memory operation behaviors within the function. When a function call statement is detected, the corresponding function summary can be directly employed to obtain the memory operation behaviors to improve the detection efficiency.
	
	The function summary is a simple description of the function features and constraint information, which is often employed for static analysis. The function summary for the memory operation behaviors is an abstract description of the memory operation behaviors within a function. This summary is generally expressed as a triple $ <v_{i}(v),  PathC  , B> $ , where $v_{i}(v)$ represents the set of variables that have ownership of the memory space; $PathC$ indicates the external conditions of the behavior; and $B$ represents the memory operation behaviors in this function. The abstract description of the memory operation behavior is $<$behavior, supplementary description of this behavior$>$ , and the description can be divided into the following situations:
	
	(a) If an object has undergone a process from memory allocation to release within a function, then the behavior is not recorded.
	
	(b) If an object is allocated memory space within a function without being freed and the object is returned or is an external object, the following behavior is recorded $action = {AlloceToExTern, <W>}$, where $W$ represents the manner in which memory is allocated.
	
	(c) If an external object has freed the originally requested memory in the function, the behavior is recorded: $action = {ExTernToFree, <F, flag>}$, where $F$ is the memory release method and flag indicates whether the memory release behavior has occurred.
	
	(d) If the final state of an object within a function is a memory leak, then the behavior is not logged. However, a memory leak defect is reported;
	
	(e) If the behavior of an object within a function is unclear, the behavior is recorded: $action = <UnKonw, Null>$;
	
	$ <v_{i}(v), PathC , B> $ is a description of the memory operation behavior across functions for a memory space. For a function, there may be memory behavior operations in multiple memory spaces. Therefore, its function summary is primarily composed of unions of $ <v_{i}(v), PathC , B> $ . Therefore, the final description of the function summary is expressed as follows:
	
	\begin{equation}
	\begin{split}
			\operatorname{abstract}(F u n c)= & <v_{1}(v),  { PathC }, { action }>\cup \\ & \ldots \cup<v_{n}(v), { PathC },  { action }>
	\end{split}
	\end{equation}
	
	Due to the overloaded mechanism in C++, a function that uses only the function name cannot be uniquely identified. Therefore, $Func$ records the file name, class name, and its name, as shown in Equation (4):
	
	\begin{equation}
	 {Func}=< {fileName,}  {className, funcName}>
	\end{equation}
	
	\subsubsection{FUNCTION SUMMARY GENERATION}
	
	To obtain the corresponding summary information, a Function Call Graph (FCG) of the source code to be detected should be generated and checked to determine whether a ring exists. The existence of the ring is essentially an incorrect call relationship, which indicates the possibility of an infinite loop in the code, and the defect may be reported. We adopt a bottom-up traversal strategy for the FCG to collect the leaf nodes in the graph (these nodes do not need to rely on other functions for their summary information) and generate a CFG. The data flow analysis is performed on the nodes on the CFG, and the nodes information with the memory operation behaviors is abstracted to obtain the function summary. The function summary generation algorithm is designed as Algorithm 2.
	
	\begin{figure}[!t]
		\label{SummaryGeneration ($CFG$)}
		\renewcommand{\algorithmicrequire}{\textbf{Input:}}
		\renewcommand{\algorithmicensure}{\textbf{Output:}}
		\removelatexerror
		\begin{algorithm}[H]
			\caption{SummaryGeneration ($CFG$)}
			\begin{algorithmic}[1]
				\begin{small}
					\renewcommand{\algorithmicrequire}{\textbf{Input:}}
					\renewcommand{\algorithmicensure}{\textbf{Output:}}
					\REQUIRE $CFG$
					\ENSURE   $summaryFunc$
					\FOR {each $curNode$ in $CFG$}
					\STATE $preNodes = getPreSet(curNode)$
					\STATE $nr = null$
					\WHILE {$preN$ in $preNodes$}
					\STATE $nr =  join(preN,nr)$ 
					\ENDWHILE
					\STATE  $np = trans(nr, curNode)$
					\IF{$memory  \quad  operation$ on $np$}
					\STATE $oper = getMemoryOper(np)$
					\STATE $update(summaryFunc)$
					\ENDIF
					\ENDFOR
					\RETURN $summaryFunc$
				\end{small}
			\end{algorithmic}
		\end{algorithm}
	\end{figure}
	
	The time complexity of Algorithm 2 is $O(n*count)$ , where $n$ denotes the number of nodes on the CFG, and $count$ denotes the maximum number of precursor nodes.
	
	\subsubsection{FUNCTION SUMMARY UPDATE}
	
	Algorithm 2 pertains to the case without a function call. In the actual detection process, however, the functions are interdependent. To obtain summary information about all functions in the program to be detected, obtaining the summary information of the bottom nodes on the FCG, and then performing a bottom-up update of the summary layer by layer are necessary. The error is handled when the FCG has rings. The function summary update algorithm (Algorithm 3) addresses the case in which no rings exist on the FCG.
	
	\begin{figure}[!t]
		\label{SummaryUpdate ($FCG$)}
		\renewcommand{\algorithmicrequire}{\textbf{Input:}}
		\renewcommand{\algorithmicensure}{\textbf{Output:}}
		\removelatexerror
		\begin{algorithm}[H]
			\caption{SummaryUpdate ($FCG$)}
			\begin{algorithmic}[1]
				\begin{small}
					\renewcommand{\algorithmicrequire}{\textbf{Input:}}
					\renewcommand{\algorithmicensure}{\textbf{Output:}}
					\REQUIRE $FCG$
					\ENSURE   $allSummaryFunc$
					\STATE $genList = getLeafNode(FCG)$
					\WHILE {$!(genList.isempty())$}
					\STATE $genFunc = getFunc(genList)$
					\IF{$isUpdateSubFunc(genFunc)$}	
					\STATE $summaryFunc = getSummaryFunc(genFunc)$
					\STATE $allSummaryFunc.add(summaryFunc)$
					\ELSE
					\STATE $genList.add(genFunc)$
					\ENDIF
					\FOR {function $f$ calls $genFunc$ on $FCG$}
					\IF{$f$ is not in $genList$}
					\STATE $genList.add(f)$
					\ENDIF
					\ENDFOR
					\ENDWHILE
					\RETURN $summaryFunc$
				\end{small}
			\end{algorithmic}
		\end{algorithm}
	\end{figure}
	
	The core idea of the algorithm is to use a singly linked list to save the functions to be updated, where we consider the header node as the current function to be updated, traverse the child nodes on the $FCG$, and check whether it has been updated. If it has been updated, then the function of the summary update is performed; if it has not been updated, the function is added to the end of the singly linked list. After the function is updated, it traverses all nodes on the FCG and adds them to the end of the singly linked list by order. The function of the summary update algorithm is described as Algorithm 3.
	
	The time complexity of Algorithm 3 is divided into two parts: a summary generation algorithm for a single function $O(n*count)$ and the number $countFunc$ of functions on the $FCG$, where $n$ represents the maximum number of precursor nodes of the CFG. Thus, the time complexity of Algorithm 3 is $O(n*count*countFunc)$ .
	
	\subsection{MEMORY LEAK DETECTION}
	
	\subsubsection{GENERAL MEMORY LEAK DETECTION}
	
	General memory leak detection consists of memory leak detection in a function with no call relationships or multiple call relationships. The core is the establishment of a state machine based on memory operation behaviors and control of the changes of the state machine. The state machine is established by obtaining the memory allocation statement and is directly changed from the initial “START” to “ALLOCED”. The fuzzy matching algorithm is combined with the function summary information to analyze the subsequent statements to control the state machine. For a function statement analysis, statements in functions that may have memory operations are handled, such as assignments, memory allocation, memory free, returns, function calls, and if-else branches. The general memory leak detection algorithm is described as Algorithm 4.
	
	\begin{figure}[!t]
		\label{GeneralCheck ($program$)}
		\renewcommand{\algorithmicrequire}{\textbf{Input:}}
		\renewcommand{\algorithmicensure}{\textbf{Output:}}
		\removelatexerror
		\begin{algorithm}[H]
			\caption{GeneralCheck ($program$)}
			\begin{algorithmic}[1]
				\begin{small}
					\renewcommand{\algorithmicrequire}{\textbf{Input:}}
					\renewcommand{\algorithmicensure}{\textbf{Output:}}
					\REQUIRE $program$
					\ENSURE   $memoryLeakReport$
					\STATE $genSummary(program)$
					\FOR {each $func$ in $program$}					
					\IF{$Visit(func)$}
					\STATE continue
					\ENDIF
					\STATE $nodeCFG = genCFG(func)$
					\STATE $BFS(nodeCFG)$
					\ENDFOR
					\RETURN $memoryLeakReport$
				\end{small}
			\end{algorithmic}
			\begin{algorithmic}[2]
			\begin{small}
				\renewcommand{\algorithmicrequire}{\textbf{Input:}}
				\renewcommand{\algorithmicensure}{\textbf{Output:}}
				\STATE $BFS(nodeCFG)$
				\REQUIRE $nodeCFG$
				\ENSURE   $memoryLeakReport$
				\WHILE{each $node$ in $CFG$ is not null}
				\IF{$node$ is operation about memory}
				\STATE $update(StateMachine)$
				\IF{$StateMachine$ is $"ERROR"$}
				\STATE $memoryLeakReport.add()$
				\ENDIF
				\ELSIF{$node$ is function call}
				\STATE $getSummary(func)$
				\STATE $update(StateMachine)$
				\IF{$StateMachine$ is $"ERROR"$}
				\STATE $memoryLeakReport.add()$
				\ENDIF
				\ELSIF{$node$ is conditional branch}
				\FOR{each $nodeBranch$ in $conditional branch$}
				\STATE BFS($nodeBranch$)
				\ENDFOR
				\ENDIF
				\ENDWHILE
				\RETURN $memoryLeakReport$
			\end{small}
		\end{algorithmic}
		\end{algorithm}
	\end{figure}
	
	The time complexity of Algorithm 4 is $O(n^3*countFunc)$ , where $countFunc$ represents the number of functions on the $FCG$ and $n$ represents the maximum number of precursor nodes of the $CFG$. Usually, $n$ is very small.
	
	\subsubsection{SPECIAL MEMORY LEAK DETECTION}
	
	Special memory leak detection is a special rule detection for C++ \cite{zhou2016design}, which is the memory leak detection of class members summarized in the memory leak defect mode.
	
	For the first type, the memory leak is caused by the inconsistency of the constructor and the destructor; first, we need to obtain all class definitions. For each class, the constructor is checked for the memory allocation behaviors of the class members. If it exists, an abstract description of the corresponding memory allocation behavior is generated. Second, the destructor is checked for the memory release behaviors of the class members; if it exists, an abstract description of the corresponding memory release behavior is generated. Last, the abstract description of the memory release behavior is matched against the abstract description of the memory allocation behavior, and the mismatch is reported.
	
	For the second type, the destructor of the parent class with the subclass is not virtual. First, we need to obtain the parent-child relationship of all classes in the detection program. Second, the destructor of each parent class is checked; if it does not exist or it is not virtual, an error is reported.
	
	For the third type, the copy constructor or assignment overloaded function is not properly defined. First, we need to obtain all class definitions in the test program. Second, we need to determine whether a class member variable points to the memory space in the class and whether the class has a custom copy constructor or assignment overload function. Last, we have to check whether an improper object copy mode (shallow copy) is used to construct the object.
	
	A special memory leak detection algorithm is described as Algorithm 5.
	
	\begin{figure}[!t]
		\label{SpecialCheck ($program$)}
		\renewcommand{\algorithmicrequire}{\textbf{Input:}}
		\renewcommand{\algorithmicensure}{\textbf{Output:}}
		\removelatexerror
		\begin{algorithm}[H]
			\caption{SpecialCheck ($program$)}
			\begin{algorithmic}[1]
				\begin{small}
					\renewcommand{\algorithmicrequire}{\textbf{Input:}}
					\renewcommand{\algorithmicensure}{\textbf{Output:}}
					\REQUIRE $program$
					\ENSURE   $memoryLeakReport$
					\STATE $classInfo = getClassInformation(program)$
					\STATE $relationMap = getRelation(classInfo)$
					\FOR {$parentClass$ in $relationMap$}
					\IF{$!checkVirtual(parentClass)$}
					\STATE $memoryLeakReport.add()$
					\ENDIF
					\ENDFOR
					\FOR{each $class$ in $classInfo$}
					\STATE $allocInfo = getAlloc(class)$
					\STATE $freeInfo = getFree(class)$
					\IF{$!match(allocInfo,freeInfo)$}
					\STATE $memoryLeakReport.add()$
					\ENDIF
					\STATE $cpInfo = checkCopyorAssignment(class)$
					\IF{($cpInfo$ is $error$) $||$ ($cpInfo==NULL$ \&\&
						default copy constructor or assignment function)
						}
					\STATE $memoryLeakReport.add()$
					\ENDIF
					\ENDFOR					
					\RETURN $summaryFunc$
				\end{small}
			\end{algorithmic}
		\end{algorithm}
	\end{figure}	
	
	The time complexity of Algorithm 5 is $O(num*n)$ , where $num$ represents the number of classes defined in the program, and $n$ represents the detection time for a single class.
	
	\section{EXPERIMENTS}
	
	\subsection{EXPERIMENTAL ENVIRONMENT}
	
	Based on the designed memory leak detection method, we implement a static detection system named ZkCheck. The experimental environment of this paper is shown in Table \ref{EXPERIMENTAL ENVIROMENT}.

	\begin{table}[!t]
		\renewcommand{\arraystretch}{1.3}
		\caption{EXPERIMENTAL ENVIROMENT}
		\label{EXPERIMENTAL ENVIROMENT}
		\centering
		\begin{tabular}{p{4.0cm} p{4.0cm}}
			\hline\hline
			OS                                                 & ubuntu14.04, windows                            \\
			memory                                             & 4G                                              \\
			cpu                                                & X64 (i5)                                        \\
			Comparison Method (memory leak   detection tool)   & Clang Static Analyzer, Sparrow, Clouseau        \\
			Dataset (open source C\textbackslash{}C++ project) & gcc, ammp, bash, mesa, cluster, openCV, bitcoin	 \\
			\hline\hline                                  
		\end{tabular}
	\end{table}

	\subsection{EVALUATION INDICATORS}
	
	To evaluate the detection efficiency of static detection tools, the false positive rate, false negative rate and detection time are commonly employed as evaluation indicators. We represent the total number of defects detected by static detection as $C$; the program should have the total number of defects, which is denoted as $a c t_{-} C$; and the number of false positives is $FC$. The definitions of false positive rate ($FPR$) and false negative rate ($FNR$) are expressed as follows:

	\begin{equation}
		\mathrm{FNR}=\frac{\left|a c t_{-} C-C\right|}{a c t_{-} C}
	\end{equation}

	The calculation formula of $FPR$ is expressed as follows:
	
	\begin{equation}
		\mathrm{FPR}=\frac{F C}{C}
	\end{equation}

	In Equation (5), since obtaining the actual number of defects in the source program to be detected $a c t_{-} C$ is difficult, the false positive rate is the main focus of the experimental results in this paper.

	\subsection{EXPERIMENTS AND RESULT ANALYSIS}
	
	\subsubsection{VALIDATION EXPERIMENT}
	
	\begin{table}[!t]
	\renewcommand{\arraystretch}{1.3}
	\caption{RESULTS OF VALIDATION EXPERIMENT}
	\label{RESULTS OF VALIDATION EXPERIMENT}
	\centering
		\begin{tabular}{lllll}
			\hline\hline 
			program & size(KLOC) & Times(secs) & Bug   Count & False   Count \\
			\hline
			gcc     & 230.4      & 213.1       & 36          & 6             \\
			ammp    & 13.4       & 10.4        & 23          & 5             \\
			bash    & 100.0      & 90.1        & 16          & 3             \\
			mesa    & 61.3       & 48.6        & 9           & 8             \\
			cluster & 10.7       & 9.5         & 12          & 4             \\
			openCV  & 794.6      & 756.8       & 74          & 11            \\
			bitcoin & 94.4       & 78.7        & 22          & 7             \\
			Total   & 1304.8     & 1257.9      & 192         & 44  \\
			\hline\hline         
		\end{tabular}
	\end{table}
	
	Use the proposed algorithm to detect the open source projects in Table \ref{EXPERIMENTAL ENVIROMENT}, and test the defects to obtain the number of false positives. The experimental results are shown in Table. \ref{RESULTS OF VALIDATION EXPERIMENT} .
	
	The experimental results show that the detection speed is approximately 1.1Kloc/s, and the false positive rate is approximately 23.8\%. By the analysis of the experimental results, the feasibility of the static memory leak detection method based on the defect mode is further verified.
	
	\begin{figure}[!t]
		\centering
		\includegraphics[width=2.5in]{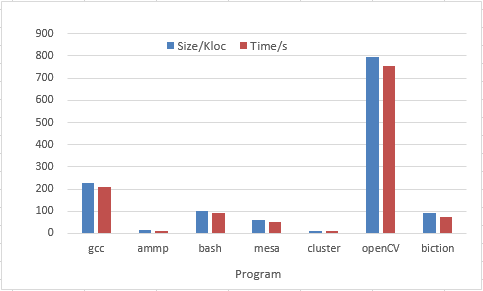}
		\DeclareGraphicsExtensions.
		\caption{Comparison of code quantity and detection time}
		\label{Comparison of code quantity and detection time}
	\end{figure}

	\begin{figure}[!t]
		\centering
		\includegraphics[width=2.5in]{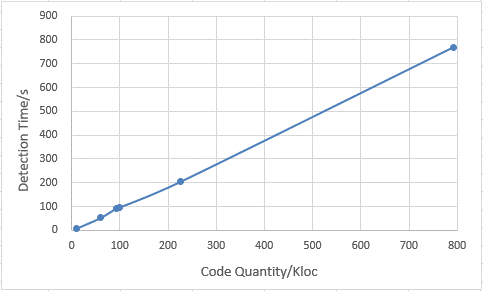}
		\DeclareGraphicsExtensions.
		\caption{Relation between code quantity and detection time}
		\label{Relation between code quantity and detection time}
	\end{figure}

	\begin{figure}[!t]
		\centering
		\includegraphics[width=2.5in]{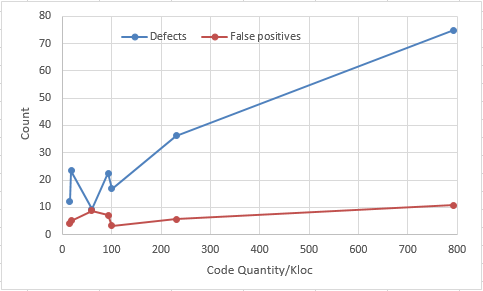}
		\DeclareGraphicsExtensions.
		\caption{Comparison of code quantity with the number of reported defects and the number of false positives}
		\label{Comparison of code quantity with the number of reported defects and the number of false positives}
	\end{figure}
	
	The experimental results obtained in Table \ref{RESULTS OF VALIDATION EXPERIMENT} are analyzed, and a certain relationship between the code quantity in the open source project and the time required for detection is identified. The experimental results are shown in Fig. \ref{Comparison of code quantity and detection time} and Fig. \ref{Relation between code quantity and detection time}. In Fig. \ref{Comparison of code quantity and detection time}, the dark color represents the number of code lines in the open source project, and the light color represents the detection time required by the open source project. By the analysis of Fig. \ref{Comparison of code quantity and detection time}, a positive correlation between the number of code lines in the open source project and the required detection time is observed. The larger is the code of open source project, the longer is the time required to detect the code.
	
	As shown in Fig. \ref{Relation between code quantity and detection time}, a certain positive correlation between the code quantity and the detection time is observed. The relation between the code quantity and the detection time is approximately a straight line, and the slope of this line is 0.9. The time required to detect 1000 lines of code is approximately 0.9 s, which can be converted to 1111.1 lines per second, which is similar to the experiment result of 2,1137 lines per second. Further analysis of the experimental results in Table \ref{RESULTS OF VALIDATION EXPERIMENT} and combining Fig. \ref{An example of incorrect assignment overloaded function} with Fig. \ref{An example of constructor destructor memory leak} concludes a positive correlation between the detection time and the code quantity in the detected projects.
	
	The results of comparing the code quantity of the detected project with the number of reported defects and the number of false positives are shown in Fig. \ref{Comparison of code quantity with the number of reported defects and the number of false positives}. The results indicate no correlation between the code quantity and the number of reported defects or the number of false positives. As the amount of code increases, the number of reported defects may increase or decrease, which is primarily determined by the project source code. Comparing the open source project $ammp$ with mesa, $ammp$ has substantially less code but a larger number of reported errors than $mesa$. By analyzing the open source code of $ammp$ and $mesa$, the reason for the large number of defects reported by $ammp$ is the existence of multiple memory applications in ammp. However, only the last memory application is released prior to function return.
	
	\subsubsection{CONTRAST EXPERIMENT}
	
	The compared detection tools, the detection time and the number of reported defects and false reports are shown in Table \ref{RESULTS OF EXPERIMENT 2}. For the contrast tools selected in the experiment, the specific implementation methods are listed as follows: Clouseau uses flow-sensitive and context-sensitive analysis techniques based on ownership models; Sparrow employs a path-insensitive analysis technique and a function summary to handle the function call expansion; and CSA uses both a compliance analysis and the constraint solution. The detected open source code is primarily expressed in C language. 
	
	\begin{table}[!t]
	\renewcommand{\arraystretch}{1.3}
	\caption{RESULTS OF EXPERIMENT 2}
	\label{RESULTS OF EXPERIMENT 2}
	\centering
		\begin{tabular}{p{2cm}p{1cm}p{1cm}p{2cm}}
			\hline\hline
			Leak   Detector             & Speed (Loc/sec) & Bug   Count & False   Positive Rate (\%) \\
			\hline
			Clang Static Analyzer (CSA) & 400            & 81          & 27                         \\
			Sparrow                     & 720            & 69          & 19                         \\
			Clouseau                    & 500            & 92          & 41                         \\
			ZkCheck                     & 1137           & 102         & 24      \\
			\hline\hline     
		\end{tabular}
	\end{table}

\begin{figure}[!t]
		\centering
		\includegraphics[width=2.5in]{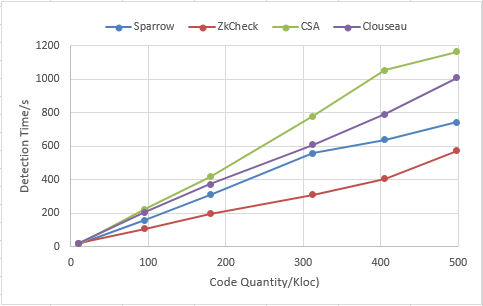}
		\DeclareGraphicsExtensions.
		\caption{Comparison of detection tools for detection time}
		\label{Comparison of detection tools for detection time}
	\end{figure}

\begin{figure}[!t]
		\centering
		\includegraphics[width=2.5in]{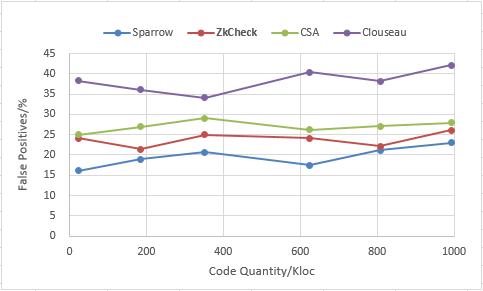}
		\DeclareGraphicsExtensions.
		\caption{Comparison of detection tools for false positive rate}
		\label{Comparison of detection tools for false positive rate}
	\end{figure}

	Fig. \ref{Comparison of detection tools for detection time} and Fig. \ref{Comparison of detection tools for false positive rate} show the comparison results of the detection tools for detection time and false negative rate, respectively. An analysis of the experimental results reveals that our method has the fastest detection speed; is superior to Clouseau, CSA, Sparrow and other tools; and has a stable detection speed. The method will not cause an explosion of the detection time due to a massive increase in code. In terms of the false negatives rate, our method is lower than Clouseau, similar to CSA, and higher than Sparrow. In terms of the number of reported defects, our method is superior to CSA and Sparrow.
	
	Analyzing the detection methods employed by these four tools, for ZkCheck and Sparrow, the function summary is used to reduce the repeated spread detection problems in the function call points, which improves the detection speed and reduces the detection time. However, ZkCheck differs from Sparrow because ZkCheck uses control flow to analyze the branch paths in the detected program, whereas Sparrow uses path-insensitive analysis, which omits path-related defects and a relatively low number of defects.
	
	Regarding the detection tool CSA, it employs the detection methods of symbolic analysis and constraint solving and has high precision in interval operations, pointer analysis and variable values. However, this tool only applies to detection within functions and cannot detect the memory operation behaviors across functions. Therefore, CSA is not superior to ZkChek in terms of the detection time and false positive rate.
	
	The detection tool Clouseau uses the method of flow sensitivity and context sensitivity analysis based on memory ownership. This method establishes the memory ownership management model, in which each object is owned by a unique pointer, and the ownership can be transferred between two pointers or deleted. Due to the lack of precision in pointer operation and the insensitivity to the path, this method has a general detection efficiency compared with other methods and is inferior with regards to the detection time and accuracy.
	
	The results of Experiment 2 show that ZkCheck has a relatively satisfactory comprehensive efficiency compared with other detection tools and improves its detection speed based on ensuring a certain low false positive rate.
	
	\begin{figure*}[!t]
		\centering
		\includegraphics[width=20cm]{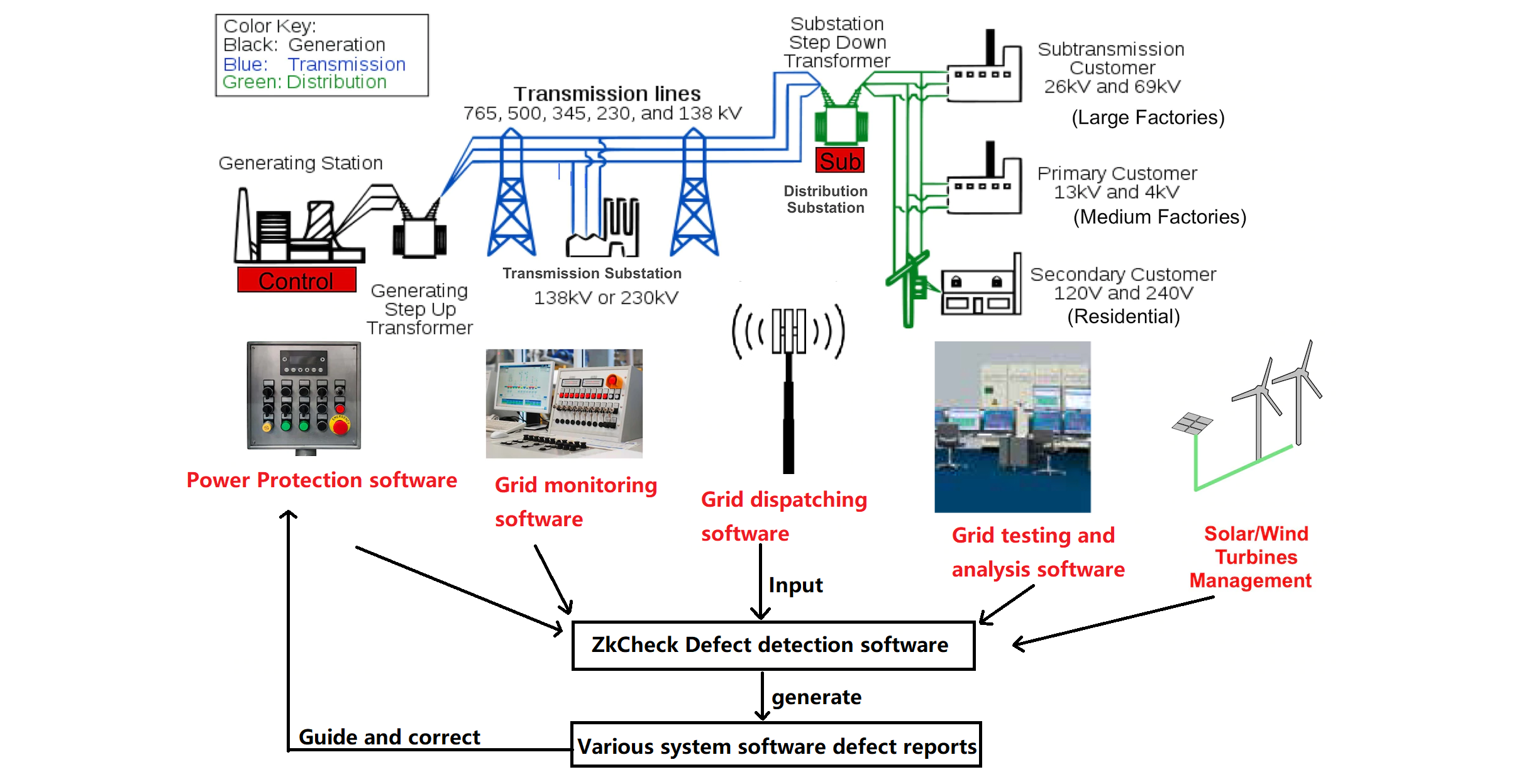}
		\DeclareGraphicsExtensions.
		\caption{Application of ZkCheck Algorithm in Industrial Smart Grid}
		\label{saving-power-smart-grid-iot-fig1}
	\end{figure*}

	\begin{figure}[!t]
	\centering
	\includegraphics[width=2.5in]{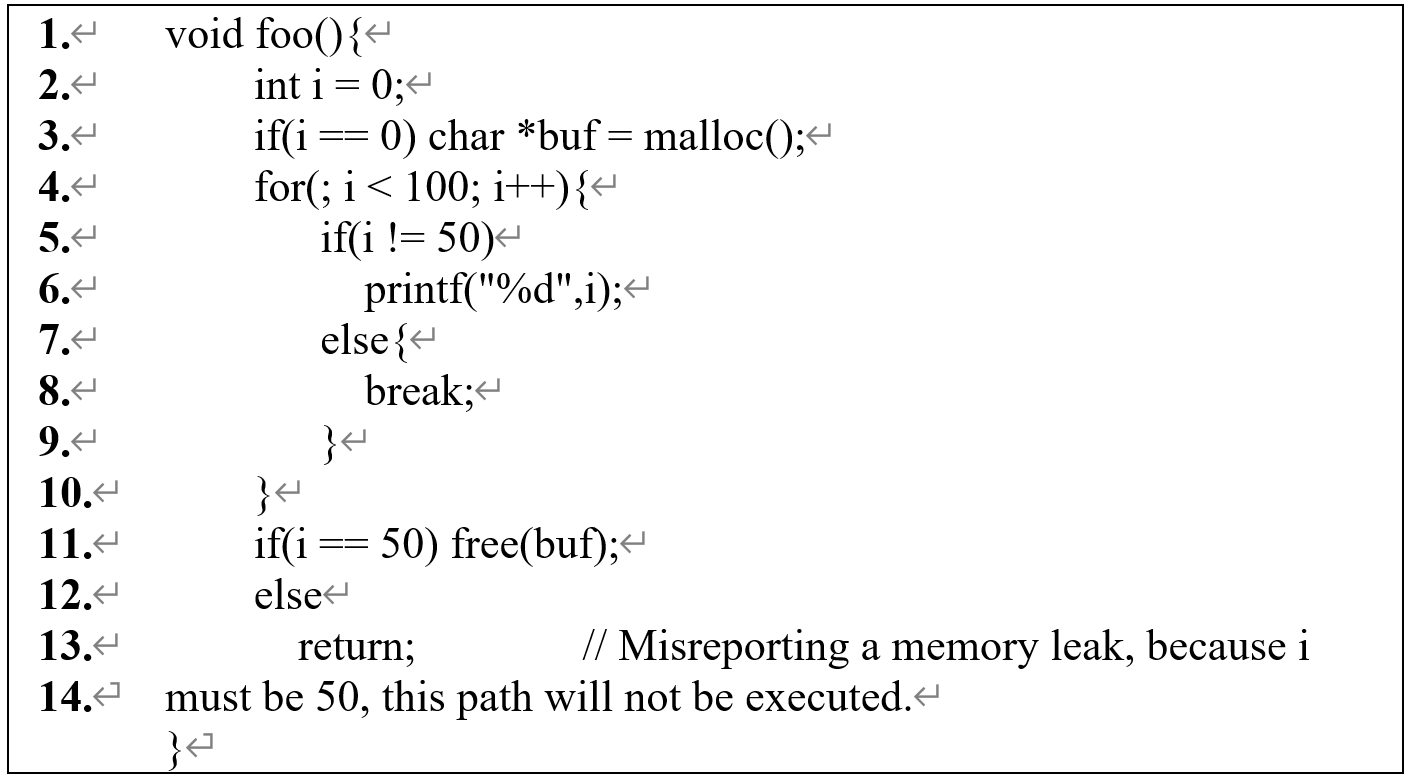}
	\DeclareGraphicsExtensions.
	\caption{Analysis of false positivies caused by insufficient intervel airthmetic}
	\label{ANALYSIS OF FALSE POSITIVES CAUSED BY INSUFFICIENT INTERVAL ARITHMETIC}
	\end{figure}
	
	\subsubsection{CAUSE ANALYSIS OF FALSE POSITIVES AND FALSE NEGATIVES}
	
	The analysis of false reported defects indicate that the causes of false positives in the memory leak detection methods proposed in this paper include the following aspects:
	
	(1) The lack of interval arithmetic precision causes a lack of sensitivity analysis of paths on the CFG, which yields a false positive. As shown in Fig. \ref{ANALYSIS OF FALSE POSITIVES CAUSED BY INSUFFICIENT INTERVAL ARITHMETIC}, due to the lack of analysis precision in the interval operation of $i$ in the loop, line 12 is considered to be a reachable path. Thus, line 13 may be incorrectly reported to have a memory leak defect.
	
	(2) Insufficient handling of overloaded functions of C++ language. Due to the overloaded mechanism of C++ language, the function called at the actual runtime without executing the program cannot be obtained, which produces false positives caused by the function summary that differ at the function call point and the actual runtime.
	
	(3) For function recursion, insufficient analysis of memory operation behaviors in circulation can easily cause false positives.

	\subsection{Industrial applications}
	
	The above experiment show that ZkCheck has a relatively satisfactory comprehensive efficiency compared with other detection tools and improves its detection speed based on ensuring a certain low false positive rate. The memory leak detection algorithm we proposed can be widely used in the detection of major actual smart grid software.
	
	In the operation of large-scale real-time software such as smart grid software, memory leaks are very common in memory defects, usually caused by improper management of dynamic memory in program implementation. When the dynamically allocated memory in the program is not released in time, there will be leakage, which will cause the memory space to be consumed and cannot be recycled and reused, resulting in a reduction in memory leakage resources, and ultimately reducing program performance, resulting in the overall smart grid system collapse. At the same time, memory leak has a certain degree of concealment, it is easy to be ignored, especially for the language of C / C ++ such a significant memory management, the expansion of the expansion scale and the programmer's irregular programming, memory leaks The situation is more common. This is very common in large-scale smart grid software.
	
	Our detection algorithm ZkCheck is a static detection algorithm, which can detect errors in the smart grid software before it enters the network and runs. At the same time, experiments show that our algorithm performs better than other detection algorithms in terms of detection speed and accuracy. Our algorithm is compatible with various softwares. As shown in the Fig \ref{saving-power-smart-grid-iot-fig1}, in all kinds of software during the operation of the smart grid, our algorithm can detect the defects and generate the corresponding detection report to further guide the further modification of the smart grid software.
	
	In other industrial fields, our detection algorithms are also versatile, accurate and efficient, and can be extended to software defect detection in other fields.

	\section{Conclusion}
	Because the smart grid software is large in scale and needs to run in real time, there is an urgent need for a memory leak detection algorithm that can find software defects before the software enters the network for operation. In this paper, we present a static method for memory leak detection based on a defect mode. This method statically detects leaks in C/C++ programs. We preprocess the source code, perform lexical analysis, and convert it into a two-way linked list with language keywords as the unit. We add corresponding symbol table information to the two-way linked list. By matching the two-way linked list with the corresponding defect mode to control the change in the defect state machine, memory leak defect detection can be realized. By comparing the existing detection tools, our experimental results show that the static detection method of memory leak based on the defect mode  has a good perform in discovering potential memory leak defects in program source code, which can effectively improve the correctness of developed software. The memory leak detection algorithm we proposed can be widely used in the detection of major actual smart grid software.
	
	The following aspects can be explored in subsequent research:
	
	1. Provide corresponding solutions for the detected defects.
	
	2. Path-sensitive processing. Path explosions may occur due to excessive conditional statements, loop statements and function recursions in the encoding. Therefore, effective analysis of the convergence of the path nodes based on ensuring efficiency is the focus of future research.
	
	3. Improve the accuracy of detection. For static detection, the inaccuracy of interval arithmetic in the detection process may cause a large number of false negatives or false positives. Constant changing of the state machine may generate path analysis errors due to the continuous amplification of inaccuracy. The difficulty of interval arithmetic pertains to addressing floating point precision, the loop process and function recursion. Methods for improving the calculation accuracy in the detection will be critical to future research.

	\ifCLASSOPTIONcaptionsoff
	\newpage
	\fi
	
	\bibliographystyle{IEEEtran}
	\bibliography{mylib}
	
	\begin{IEEEbiography}[{\includegraphics[width=1in,height=1.25in,clip,keepaspectratio]{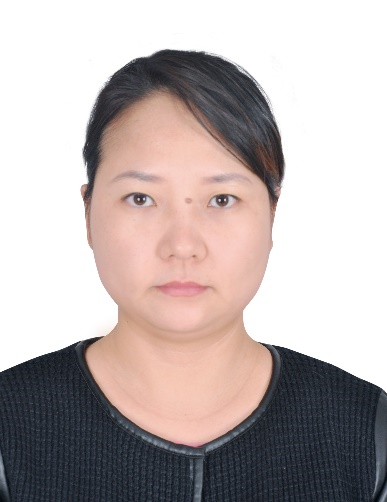}}]{Ling Yuan}
		is currently an associate professor in School of Computer Science and Technology, HUST, Wuhan, P.R. China. She received her Ph.D. degree in the Department of Computer Science from National University of Singapore in 2008. Her research interest includes data processing, and software engineering.
	\end{IEEEbiography}
	
	\begin{IEEEbiography}[{\includegraphics[width=1in,height=1.25in,clip,keepaspectratio]{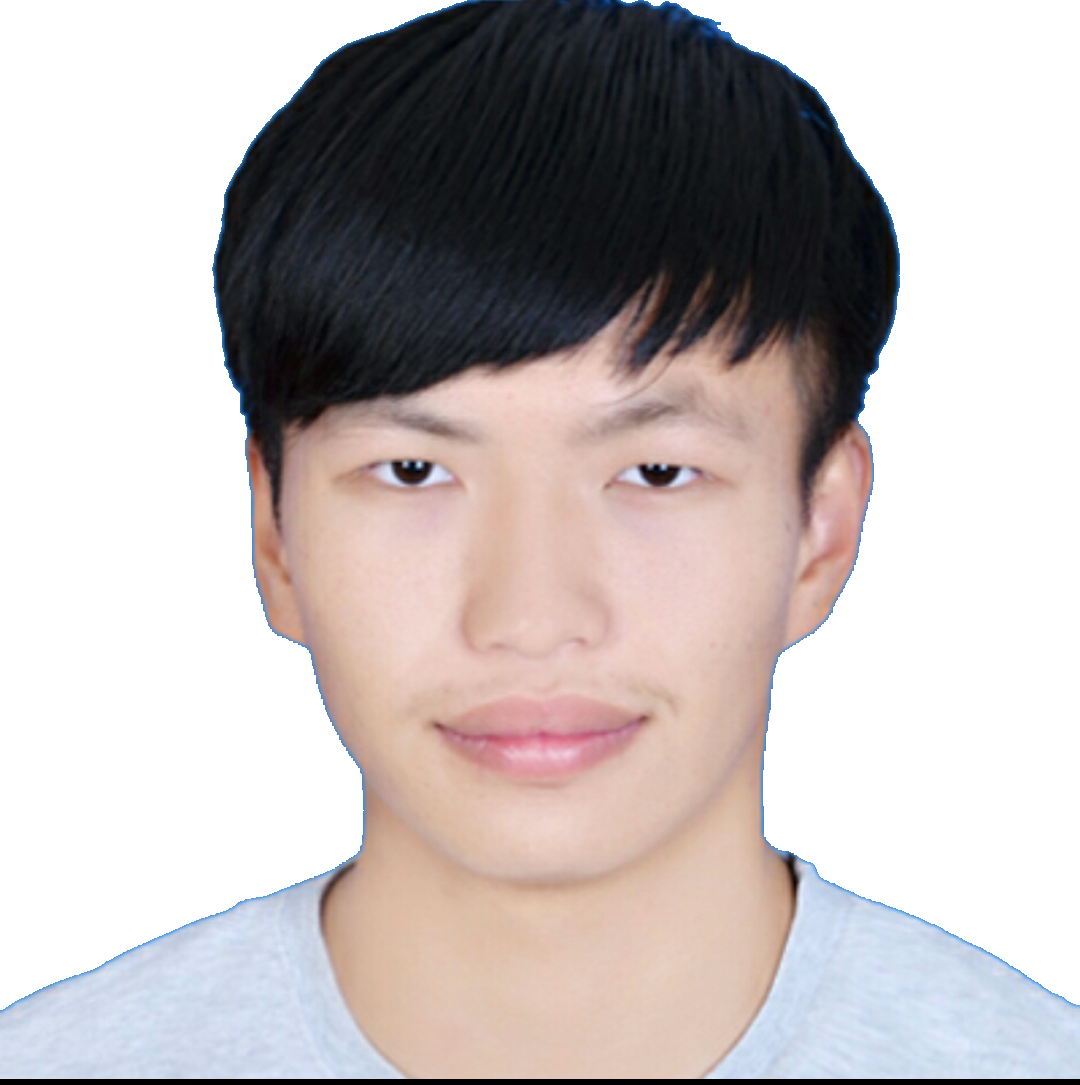}}]{Siyuan Zhou}
		received the B.E. degree in Software engineering from Dalian University of Technology, Liaoning, China, in 2019. He is currently pursuing the master's degree in computer software and theory in School of Computer Science and Technology, HUST, Wuhan, China.
		His current research interests include data processing, and software engineering.
	\end{IEEEbiography}
	
	\begin{IEEEbiography}[{\includegraphics[width=1in,height=1.25in,clip,keepaspectratio]{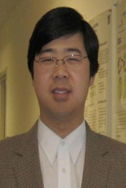}}]{Neal N. Xiong}
		is current a Professor at the Department of Computer Science, Southwestern Oklahoma State University, USA. He received his both PhD degrees in Wuhan University (about software engineering), and Japan Advanced Institute of Science and Technology (about dependable networks), respectively. Before he attends Colorado Technical University, he worked in Wentworth Technology Institution, Georgia State University for many years. His research interests include Cloud Computing, Security and Dependability, Parallel and Distributed Computing, Networks, and Optimization Theory. He is a Senior member of IEEE Computer Society.	
	\end{IEEEbiography}

\end{document}